\documentclass[12pt]{article}
\def\e{{\rm e}}
\def\d{{\rm d}}

\def\smhalf{  {\textstyle{1\over 2} }}
\def\lam{  \lambda  }
\def\flam{  f(\lam) }
\def\glam{  g(\lam) }
\def\cN{  {\cal N}  }
\def\cA{  {\cal A}  }
\def\cO{  {\cal O}  }
\def\cAtree{  {\cal A}_{\rm tree}  }
\def\cAdiv{  {\cal A}_{\rm div}  }
\def\tC{ \tilde{C}  }
\def\theequation{\thesection.\arabic{equation}}

\begin{document}
\bibliographystyle{bst}

\begin{flushright}
BRX-TH-590\\
BOW-PH-141\\
\end{flushright}
\vspace{30mm}

\vspace*{.3in}

\begin{center}
{\Large\bf\sf  
Regge behavior of gluon scattering amplitudes in 
$\cN=4$ SYM theory 
}
\vskip 5mm Stephen G. Naculich\footnote{Research supported in part
by the NSF under grant PHY-0456944}$^{,a}$
and Howard J.  Schnitzer\footnote{Research supported in part 
by the DOE under grant DE--FG02--92ER40706\\
{\tt \phantom{aaa} schnitzr@brandeis.edu; naculich@bowdoin.edu}\\
}$^{,b}$

\end{center}

\begin{center}
$^{a}${\em Department of Physics\\
Bowdoin College, Brunswick, ME 04011}

\vspace{.2in}

$^{b}${\em Theoretical Physics Group\\
Martin Fisher School of Physics\\
Brandeis University, Waltham, MA 02454}
\end{center}
\vskip 2mm

\begin{abstract}
It is shown that the four-gluon scattering amplitude 
for $\cN=4$ supersymmetric Yang-Mills theory in the planar limit
can be written, in both the weak- and strong-coupling limits,
as a reggeized amplitude,
with a parent trajectory and
an infinite number of daughter trajectories.
This result is not evident {\it a priori},
and relies crucially on the fact 
that the leading IR-divergence and the finite $\log^2 (s/t)$-dependent piece
of the amplitude are characterized
by the same function for all values of the coupling,
as conjectured by Bern, Dixon, and Smirnov,
and proved by Alday and Maldacena in the strong-coupling limit.
We use the Alday-Maldacena result to 
determine the exact strong-coupling Regge trajectory.

\end{abstract}

\vfil\break

%\renewcommand{\baselinestretch}{2}
%\small\normalsize

\section{Introduction and Conclusion}
\renewcommand{\theequation}{1.\arabic{equation}}
\setcounter{equation}{0}
\label{secintro}

In this note, we analyze the Regge behavior of the
four-gluon scattering amplitude for $\cN=4$ supersymmetric SU($N$) 
Yang-Mills theory
in the planar (large $N$) limit,
using the conjectured ansatz of 
Bern, Dixon, and Smirnov \cite{Bern:2005iz}
and the recent strong-coupling results of 
Alday and Maldacena \cite{Alday:2007hr}
obtained via the AdS/CFT correspondence.
(Other recent applications of this work 
include refs.~\cite{Bern:2007ct,Drummond:2007au}.
Reggeization of the gluon in nonsupersymmetric Yang-Mills 
theories \cite{Grisaru:1973vw} as well as supersymmetric 
Yang-Mills theories \cite{Grisaru:1981ra}
has long been a subject of interest.)
The Regge limit corresponds to 
center-of-mass energy squared $u \to \infty$ 
with fixed spacelike momentum transfer $s<0$, 
where
$s =  (k_1 + k_2)^2$, 
$t =  (k_1 + k_4)^2 $, and 
$u =  (k_1 + k_3)^2$  are Mandelstam variables
%(with metric $(+,-,-,-)$) 
obeying $s+t+u=0$.
We show that in the Regge limit the 
color-ordered four-gluon amplitude approaches
\begin{equation}
\label{eq:bhvr}
\cA_4 
\mathrel{\mathop{\longrightarrow}\limits_{u \to \infty}}
\beta(s)\,  \left[ \left( u \over -s \right) ^{\alpha(s)} + \cdots \right]
\end{equation}
where the leading Regge trajectory has the form
\begin{equation}
\label{eq:regge}
 \alpha(s) =  1 +  { 1\over 4 \epsilon}  \,f^{(-1)} (\lam)  
            - {1 \over 4} \flam \log  \left( -s\over \mu^2\right)
            + { 1\over 2  } \, \glam
\end{equation}
and
\begin{equation}
\label{eq:residue}
\beta(s) =  ({\rm const}) \, \cAdiv^4 (s) \, \e^{\tC(\lam)} 
\end{equation}
with $\cdots$ representing an infinite sum of subleading trajectories.
The functions $\alpha(s)$ and $\beta(s)$,
like the scattering amplitude itself,
exhibit infrared divergences, 
which we regulate using dimensional regularization 
in $d = 4 - 2 \epsilon$ dimensions.
The four-dimensional 't Hooft coupling $\lam = g^2 N$ is dimensionless,
and a scale $\mu$ is introduced to allow the coupling to be 
defined away from four dimensions. 
The functions $\flam$ and $\glam$ characterize the 
IR divergence of the 
amplitude \cite{Bern:2005iz,Alday:2007hr,Collins:1989bt}:
$\flam$ is proportional to the cusp anomalous dimension 
\cite{Gubser:2002tv},
and $\glam$ is the function ${\cal{G}}_0$ defined in ref.~\cite{Bern:2005iz}.
The form of $\glam$ is dependent on the choice of scale 
$\mu$ \cite{Alday:2007hr}.
Finally $f^{(-1)}(\lam)$ is defined via
\begin{equation}
\label{eq:fminusone}
\left( \lam { \d \over \d \lam }   \right) f^{(-1)} (\lam ) = \flam
\end{equation}
and
$\cAdiv(s)$ and $\tC(\lam)$ are defined in 
eqs.~(\ref{eq:Adiv}) and (\ref{eq:BDS}).

We emphasize that the Regge behavior of $\cA_4$ that
we have demonstrated is not {\it a priori} evident 
from the results of ref.~\cite{Bern:2005iz,Alday:2007hr},
and in fact appears inconsistent with the fact that the exponent
of the amplitude (\ref{eq:BDS}) goes as $\log^2 (t/s)$,
whereas Regge behavior would seem to require $\log(t/s)$ dependence.
The Regge behavior of the amplitude (\ref{eq:bhvr})
only occurs because the function $\flam$ that 
characterizes the leading IR divergence
also multiplies the finite $\log^2 (s/t)$-dependent piece
of the amplitude, as conjectured in ref.~\cite{Bern:2005iz}.

The Regge trajectory function (\ref{eq:regge}) 
and residue (\ref{eq:residue}) 
are exact (to all orders in the coupling) in the planar limit,
depending only on the forms of the functions $\flam$ and $\glam$.
Since these functions are known in the weak-coupling \cite{Bern:2005iz}
and strong-coupling \cite{Alday:2007hr} limits, 
we may determine the exact trajectory function explicitly in both these limits.
To lowest order in $\lam$, we have
\begin{equation}
  \alpha(s) =  1 + {\lam \over 8 \pi^2}
\left[  { 1\over  \epsilon} -  \log  \left( -s\over \mu^2\right) \right]
+ \cO(\lam^2) .
\end{equation}
This is equivalent to the result found in ref.~\cite{Schnitzer:2007rn}, 
where a different regularization scheme was used 
(see also refs.~\cite{Grisaru:1973vw,Grisaru:1981ra}).
The Regge trajectory function in the strong-coupling limit is 
\begin{equation}
  \alpha(s) 
\mathrel{\mathop{\longrightarrow}\limits_{\lam \to \infty}}
 \sqrt{\lam}
\left[  { 1\over 2 \pi  \epsilon} 
      - {1 \over 4 \pi } \log  \left( -s\over \mu^2\right)
        + {(1 - \log 2) \over 4 \pi  } 
\right]
\end{equation}
where we have used the results of Alday and Maldacena \cite{Alday:2007hr}.

A linear Regge trajectory $\alpha(s) \sim \alpha' s$ 
would imply stringy behavior, with string tension $ \sim 1/\alpha'$.
But eq.~(\ref{eq:regge}) goes as $\log(-s/\mu^2)$,
rather than linearly in $s$, suggesting that we are in 
the $\alpha' = 0$ or infinite-tension limit of a string theory,
with no Regge recurrences.
This is not unexpected since $\cN=4$ super Yang-Mills theory 
is a conformal theory, without massive states \cite{Dixon}.

After this paper was typed, we became aware that similar conclusions 
were reached using different methods in ref.~\cite{Drummond:2007au}.

\section{The BDS ansatz}
\renewcommand{\theequation}{2.\arabic{equation}}
\setcounter{equation}{0}
\label{secBDS}

To derive the result (\ref{eq:bhvr}), we begin with the
conjecture of Bern, Dixon, and Smirnov (BDS) for  
the exact form of the scattering amplitude
in the planar (large $N$) limit of $\cN=4$ supersymmetric 
SU($N$) Yang-Mills theory \cite{Bern:2005iz}.
In the kinematic region where $s$ and $t$ are both spacelike ($s$, $t<0$),
the color-ordered planar four-gluon amplitude
(expressed using the notation of ref.~\cite{Alday:2007hr})
is given by 
\begin{equation}
\label{eq:BDS}
\cA_4  = 
\cAtree \, \cAdiv^2 (s) \, \cAdiv^2 (t) \,
\exp\left[ 
		{\flam  \over 8}
                \log^2 \left(  {s\over t} \right)
	     + \tC(\lam) 
     \right] .
\end{equation}
The tree amplitude is
\begin{equation}
\cAtree =  -~{4 i K \over s t}
\end{equation}
where the definition of $K$ may be found in ref.~\cite{Bern:1998ug}.
The IR divergent contributions $\cAdiv (s)$ and $\cAdiv (t)$ 
are rendered finite in 
$d = 4 - 2 \epsilon$ dimensions, 
and take the form  
\begin{equation}
\label{eq:Adiv}
\cAdiv (s) 
= \exp\left[ - { 1\over 8 \epsilon^2 } 
     f^{(-2)} \left( { \lam \mu^{2 \epsilon } \over (-s)^\epsilon} \right) 
              - { 1 \over 4  \epsilon } 
     g^{(-1)}\ \left({ \lam  \mu^{2 \epsilon} \over (-s)^\epsilon} \right) 
       \right] 
\end{equation}
where the functions $f^{(-2)}(\lam)$ and   $g^{(-1)}(\lam)$
are related to $\flam $ and   $\glam$ via
\begin{equation}
\left( \lam { \d \over \d \lam } \right)^2 f^{(-2)}(\lam ) = \flam,
\qquad 
\left( \lam { \d \over \d \lam } \right) g^{(-1)} (\lam ) = \glam.
\end{equation}
To lowest order in $\lam$, the functions in eq.~(\ref{eq:BDS}) 
are given by \cite{Bern:2005iz}
\begin{equation}
\flam =  {\lam \over 2 \pi^2}  + \cO (\lam^2), \qquad 
\glam = \cO (\lam^2), \qquad 
\tC(\lam) = {\lam \over 16 \pi^2} \left( 4 \pi^2 \over 3 \right) + \cO (\lam^2)
\end{equation}
so that the weak-coupling scattering amplitude reads 
\begin{equation}
\cA_4 =
\cAtree\, \cAdiv^2 (s) \, \cAdiv^2 (t) \,
\exp\left[ 
	   {\lam  \over 16 \pi^2}
                   \left\{ 
               \log^2 \left( {s\over t} \right) 
	     + {4 \pi^2 \over 3} \right\} + \cO(\lam^2)
     \right] 
\end{equation}
with 
\begin{equation}
\cAdiv (s) 
= \exp\left[
- {\lam \mu^{2 \epsilon}  \over 16 \pi^2 \epsilon^2 (-s)^\epsilon } 
	 + \cO (\lam^2)
	\right].
\end{equation}

\section{The strong-coupling limit} 
\renewcommand{\theequation}{3.\arabic{equation}}
\setcounter{equation}{0}
\label{secAM}

Alday and Maldacena subsequently computed the planar four-point 
amplitude at strong coupling using  the AdS/CFT correspondence,
obtaining the result \cite{Alday:2007hr}
\begin{equation}
\cA_4 =
\cAtree\, \cAdiv^2 (s) \, \cAdiv^2 (t) \,
\exp\left[ 
           { \sqrt{\lam} \over 8 \pi }
                   \log^2 \left( {s\over t} \right)
	     + \tC (\lam) 
     \right] 
\end{equation}
with 
\begin{equation}
\cAdiv (s) 
= \exp\left[ 
       - { 1 \over 2 \pi \epsilon^2 } 
          { \sqrt{\lam \mu^{2 \epsilon } \over(- s)^\epsilon }} 
       -  { (1 - \log 2 ) \over 4 \pi \epsilon}
          { \sqrt{\lam \mu^{2 \epsilon } \over (-s)^\epsilon }} 
       \right] 
\end{equation}
and (correcting a small error)
\begin{equation}
 \tC (\lam)
=  -{ \sqrt{\lam } \over 4 \pi } 
\left[ -1 - {\pi^2 \over 3} - 2\log 2 + (\log 2)^2 \right]
\end{equation}
which is fully consistent with the BDS conjecture (\ref{eq:BDS}), with 
\begin{equation}
\flam 
\mathrel{\mathop{\longrightarrow}\limits_{\lam \to \infty}}
{ \sqrt{\lam } \over  \pi },
\qquad
\glam 
\mathrel{\mathop{\longrightarrow}\limits_{\lam \to \infty}}
{ (1 - \log 2)  \sqrt{\lam } \over 2 \pi } .
\end{equation}
%f^{(-2)}(\lam) =        { 4\sqrt{\lam } \over  \pi },
%g^{(-1)}(\lam) = { (1 - \log 2)  \sqrt{\lam} \over   \pi }

\section{Regge limit of the BDS ansatz}
\renewcommand{\theequation}{4.\arabic{equation}}
\setcounter{equation}{0}
\label{secRegge}

The purpose of this note is to examine the four-gluon scattering
amplitude in the Regge limit of large $u$, with fixed $s<0$.
Since $s+t+u=0$, this corresponds to the limit $t \to -\infty$
of the  expression (\ref{eq:BDS}).
The large $t$ behavior of $\cAtree$ is given by
\begin{equation}
\cAtree 
\mathrel{\mathop{\longrightarrow}\limits_{t \to -\infty}}
{\rm (const)~}  {t \over s}.
\end{equation}
To extract the behavior of $\cAdiv (t)$ as $t \to -\infty$,
we expand in $\epsilon$
\begin{eqnarray}
f^{(-2)} \left( { \lam \mu^{2 \epsilon } \over (-t)^\epsilon} \right) 
&=&
f^{(-2)} \left(  \lam \right) 
+  \epsilon \,f^{(-1)} (\lam)  \log \left( \mu^2 \over -t \right)  
+  \smhalf \epsilon^2\, f(\lam)  \log^2 \left( \mu^2 \over -t \right)  
+ \cO (\epsilon^3) ,
\nonumber\\
g^{(-1)}\ \left({ \lam  \mu^{2 \epsilon} \over (-t)^\epsilon} \right) 
&=& g^{(-1)}\ \left( \lam \right) 
+  \epsilon\, \glam  \log \left( \mu^2 \over -t \right)  
+ \cO (\epsilon^2) ,
\end{eqnarray}
where $ f^{(-1)} (\lam) $ is defined in eq.~(\ref{eq:fminusone}).
Hence
\begin{eqnarray}
\label{eq:first}
\cAdiv^2 (t) 
&=&
\cAdiv^2 (s) 
\exp\bigg[ { 1\over 4 \epsilon}  \,f^{(-1)} (\lam)  \log \left(t \over s \right)
           + { 1 \over 2  }                  \glam  \log \left( t \over s\right)
\nonumber\\
&& ~~~~~~~~~~~~~
-  { 1 \over 8}           f(\lam)  \log^2 \left( -t \over  \mu^2 \right)   
+  { 1 \over 8}           f(\lam)  \log^2 \left( -s \over  \mu^2 \right)   
       \bigg] .
\end{eqnarray}
Next, we rewrite the last term of eq.~(\ref{eq:BDS}) as
\begin{equation}
\label{eq:second}
\exp\left[ 
		{\flam  \over 8}
                \log^2\left( -s\over \mu^2\right) 
  	-       {\flam  \over 4}
                \log  \left( -s\over \mu^2\right)
                \log  \left( -t\over \mu^2\right) 
        +       {\flam  \over 8}
                \log^2\left( -t\over \mu^2\right) 
	     + \tC(\lam) 
     \right] .
\end{equation}
Because both the leading IR-divergence and the 
IR-finite $\log^2 (s/t)$-dependent term are controlled 
by the same function $\flam$,
we observe that the leading $\log^2(-t/\mu^2)$ dependence
cancels out of the full scattering amplitude,
so that the large-$t$ behavior is determined by the
$\log (-t/\mu^2)$ dependent terms.
Collecting the leading large-$t$ contributions to the amplitude,
we obtain 
\begin{equation}
\cA_4 
\mathrel{\mathop{\longrightarrow}\limits_{t \to -\infty}}
\beta(s)
\, \left(  t \over s \right)^{\alpha(s)}
\end{equation}
where
\begin{equation}
  \alpha(s) = 1 + { 1\over 4 \epsilon}  \,f^{(-1)} (\lam)  
            - {1 \over 4} \flam \log  \left( -s\over \mu^2\right)
            + { 1\over 2  }\, \glam
\end{equation}
and 
\begin{equation}
\beta(s) =  ({\rm const}) \, \cAdiv^4 (s)\, \e^{\tC(\lam)} .
\end{equation}
Writing the result in terms of the center-of-mass energy squared $u$, 
we obtain for the color-ordered planar four-gluon amplitude
\begin{equation}
\cA_4 
\mathrel{\mathop{\longrightarrow}\limits_{u \to \infty}}
\beta(s) \,\left( {u \over -s}  - 1   \right)^{\alpha(s)}
 = \beta(s)\,  \left[ \left( u \over -s \right) ^{\alpha(s)} + \cdots \right].
\end{equation}
The full planar four-gluon amplitude is then obtained by
summing over color-ordered amplitudes multiplied by 
the associated trace over gauge group generators.
The function $\alpha(s)$ then describes the leading Regge trajectory,
in the adjoint channel.
Any subleading terms $\cdots$ that survive would 
represent (an infinite sum of) 
daughter trajectories,  again in the adjoint channel.

\providecommand{\href}[2]{#2}\begingroup\raggedright\endgroup

\end{document}